# CMSSW Scaling Limits on Many-Core Machines

*Christopher* Jones[1]* and *Patrick* Gartung[1]

[1]Fermi National Accelerator Laboratory, Batavia, IL, USA

**Abstract.** Today the LHC offline computing relies heavily on CPU resources, despite the interest in compute accelerators, such as GPUs, for the longer term future. The number of cores per CPU socket has continued to increase steadily, reaching the levels of 64 cores (128 threads) with recent AMD EPYC processors, and 128 cores on Ampere Altra Max ARM processors. Over the course of the past decade, the CMS data processing framework, CMSSW, has been transformed from a single-threaded framework into a highly concurrent one. The first multithreaded version was brought into production by the start of the LHC Run 2 in 2015. Since then, the framework's threading efficiency has gradually been improved by adding more levels of concurrency and reducing the amount of serial code paths. The latest addition was support for concurrent Runs. In this work we review the concurrency model of the CMSSW, and measure its scalability with real CMS applications, such as simulation and reconstruction, on modern many-core machines. We show metrics such as event processing throughput and application memory usage with and without the contribution of I/O, as I/O has been the major scaling limitation for the CMS applications

## 1.Introduction

The CMS experiment evolved its offline data processing framework from being single thread to multi-thread in order to bring down the amount of memory per-core needed by the application. Using multiple threads allowed some data to be shared by multi-threads which allowed the average per-core memory usage to decrease. The application makes use of Intel's Thread Building Blocks library [1] to handle scheduling tasks to threads within a thread pool.

CMS began using the multi-threaded framework for production in 2015 as part of the start of the LHC Run 2. Since then, the application has increased the amount of concurrency it can support. Originally, the application only supported processing multiple Events concurrently (where only one Algorithm at a time was processing a given event) as well as allowing Algorithms internally to run concurrent tasks. The next step was to added the abilities to run multiple Algorithms concurrently on the same Event. The CMSSW framework's data model is composed of Runs which contain multiple LuminosityBlocks which in turn hold multiple Events. The framework now supports running multiple concurrent Runs and LuminosityBlocks as well as allowing multiple Algorithms to run within a given Run or LuminosityBlock. At this point all possible concurrency is available when processing data.

---

* Corresponding author: cdj@fnal.gov

In order to allow for high thread efficiency, CMSSW employs a novel mechanism to handle access to shared resources across algorithms. The classical mechanism for dealing with a shared resource is to place all access to the resource within critical sections. This has the disadvantage of blocking threads as they wait for their turn to get the resource. CMSSW uses a non-blocking mechanism. When an algorithm which need access to a shared resource needs to be scheduled to run, the algorithm is wrapped in a task and that task is added to a serial queue specific to that shared resource. If no task related to that serial queue is presently running, the newly added task is run. If there is a presently running task for the queue, the new task is added to the queue. Once the presently running task is completed, the last step of the task is to start running the next task in the queue. The serial task queue mechanism only requires one thread to be used for the resource, freeing up all other threads to do any other available tasks. In addition, no central scheduling is needed as either adding a task to a serial queue or the finishing of a task from a queue will cause a task in the queue to be run. Both ways of starting a task do not require any communication with any additional threads to decide what task should be run. The serial queue is thread safe and just uses atomics to deal with concurrent addition of tasks to the queue from different threads.

By using the serial queue with TBB, CMSSW is able to efficiently scheduled tasks locally on each thread without blocking any of the threads. This allows CMSSW to schedule around the serial access requirements needed by ROOT [2] I/O. Any serial access will affect the scheduling if there is insufficient other work available in the system. Such a case can happen when the serial time of a job becomes a substantial fraction of the entire job time. In this paper we will describe our work to discover the scaling limits imposed by the serial access to I/O.

## 2.Measurement Methodology

The thread scaling measurements were all performed on a single node of the US HPC Perlmutter[3] machine. The hardware being used was an AMD EPYC 7763 CPU with SSD storage. The machine consisted of 2 sockets with 64 cores per socket and each core supports 2 hardware threads. In total this provided 256 hardware threads on a single node.

The strategy applied to all measurements is as follows. First, during a measurement, enough jobs are run concurrently so that all 256 hardware threads of the node are kept busy. This avoids the possibility that the hardware will boost the CPU frequency when running a job with a low number of threads. Second, the number of events processed by a job are allows proportional to the number of threads. This tests the *weak scaling* of the system as the framework is intended to have perfect weak scaling with respect to Events. Third, the same 100 Events are repeated over and over in the input read by all jobs. As all jobs process a multiple of 100 Events, then all jobs see the exact same spread of timing for the algorithms processing an Event. Fourth, the jobs run the standard production CMS configuration for three of the standard CMS job types, which are described in the next section.

In order to disentangle the effects of I/O on the thread scaling, minor variations were applied to the job configurations. One variation replaced the algorithm used to write the ROOT files with an algorithm which would concurrently request the same data from the Event as the ROOT based algorithm but would then do nothing else with the data. This effectively removed the timing related to the serial output required by ROOT. The other variation replaced the algorithm that serially reads the Event data from ROOT as each Event is requested with an algorithm that reads the 100 unique events at the beginning of the job and then concurrently gives out an Event when the framework requests one. This removes the timing related to the serial reading of a ROOT file.

## 3.Measurements

The jobs used to measure the thread scaling are based on the workflow for generating simulated Events. The simulation workflow usually consists of three sequential jobs. The first job is to run the Event generation and detector simulation. The second overlays additional pp collisions and runs the High Level Trigger. The final job runs the full reconstruction. Each step is further described below and with the measurement results for those steps.

### 3.1.Event generation and detector simulation

The Event generation uses algorithms which simulate the proton-proton collision and creates the list of particles and four momentums of the collision products. In particular this paper generated tt-bar Events using the Pythia 8 [4] generator. This generator was chosen as it is highly used in CMS and it is thread-friendly, i.e. multiple instances of Pythia can be instantiated and each used on its own thread. The detector simulation uses the Geant4 [5] library to simulate the passage of particles through the CMS detector. The CMSSW usage of Geant 4 also allows concurrent simulation of different Events. For this configuration, the jobs were writing files with an average Event size on disk of 0.93 MB/Event.

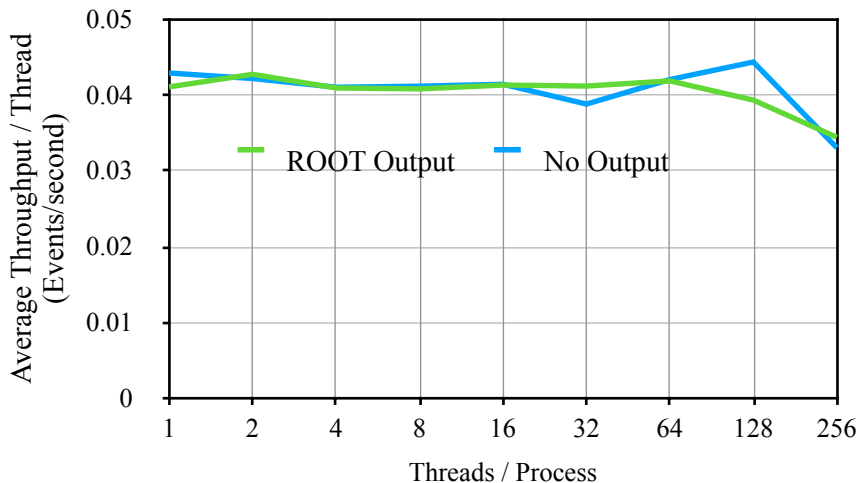

**Fig. 1.** Average event throughput per thread as a function of the number of threads used in the simulation process. Rates with and without writing output are shown.

Figure 1 shows the Event processing rate per thread for this step as a function of the number threads used by a job. For perfect thread scaling the plot would be a horizontal line. The job which writes the ROOT file scales perfectly up to about 64 threads. If the I/O is removed, the job scales perfectly up to about 128 threads. The loss of threading efficiency beyond 128 threads is not understood at this time.

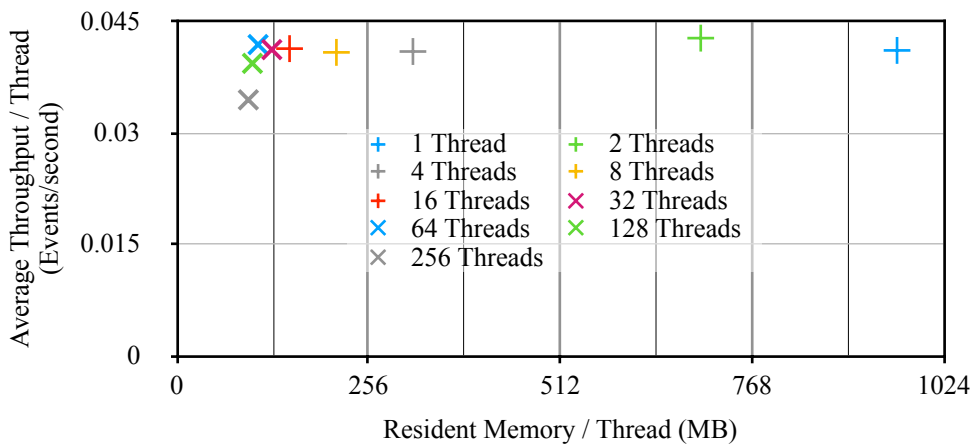

**Fig 2.** Average event throughput per thread as a function of average resident memory used per thread for simulation process. Each point represents the measurement of a process using that number of threads.

Figure 2 shows the Event processing rate per thread versus the amount of resident memory needed per thread as the number of threads used in a given job are varied from process to process. This is for the full I/O configuration of the job. The figure shows that we can have excellent thread efficiency at 64 threads and that would only require a machine have 128MB per hardware thread.

### 3.2 Overlay pp collisions and High Level Trigger

This step takes the input generated from previous simulation step, reads in additional minimum bias pp collision events and combines them into one Event. This is used to replicate the data taken by CMS as one triggered readout contains many pp collisions occurring in the detector. For this paper, we simulated the LHC environment expected for Run 3 by using 50 to 75 minimum bias Events per bunch crossing with 12 bunch crossings per input simulated tt-bar Event (where the tt-bar event is the trigger that would cause the readout of the detector). After the combining of the Events into one, all the algorithms used in the High Level Trigger are run on the unified Event. The simulated Event data is then transformed into the same representation used to store RAW data from the detector. This RAW data plus information about the underlying simulated Event are then written out to a ROOT file. For this configuration, the jobs were writing files with an average Event size on disk of 2.5 MB/Event.

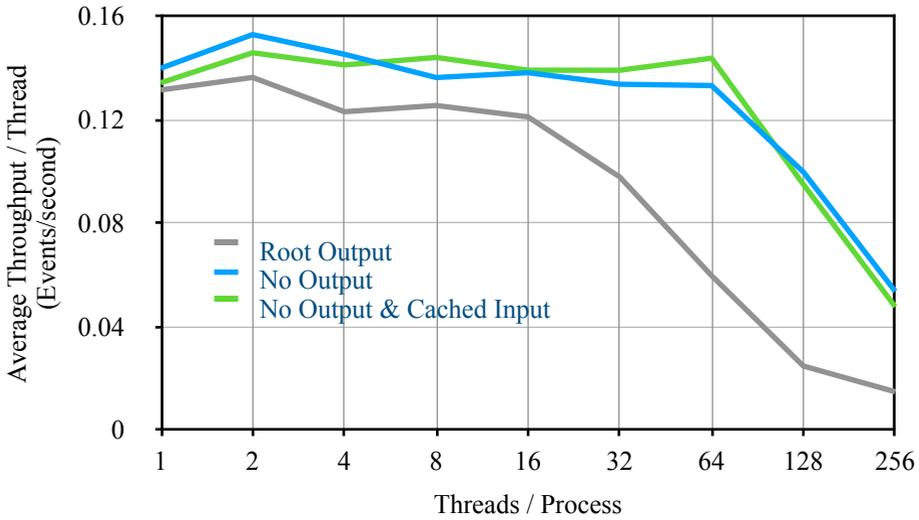

**Fig. 3.** Average event throughput per thread as a function of the number of threads used in the overlay and HLT process. Rates with and without writing output as well as without output and reading from a memory cache are shown.

Figure 3 shows the Event processing rate per thread for this step as a function of the number threads used by a job. When reading and writing ROOT files, the thread efficiency drops quickly after 16 threads. When output is removed, the application scales well out to 64 threads. Replacing input with a memory cache does not provide any further scaling benefits. The reason is beyond 64 threads, a group of algorithms which were implemented to only work serially begin to dominate the processing time. After reimplementing those algorithms the average throughput did improve (not shown) but further scaling is then hampered by the requirement to read the calibration data serially from the database.

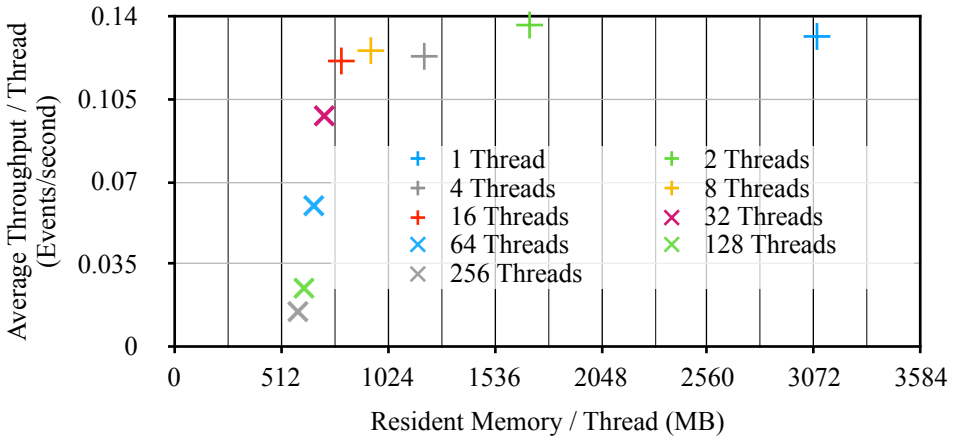

**Fig 4.** Average event throughput per thread as a function of average resident memory used per thread for overlay and HLT process. Each point represents the measurement of a process using that number of threads.

Figure 4 shows the Event processing rate per thread versus the amount of resident memory needed per thread as the number of threads used in a job are varied. This is for the full I/O configuration of the job. At least two threads are needed to have the application use less than 2GB per thread. It is possible to go below 1GB per thread and still have reasonable thread efficiency for the case of 8 or 16 threads. Beyond 16, the thread efficiency drops quickly.

### 3.3 Reconstruction

This step takes as input the RAW plus simulated data from the previous step and runs algorithms which ultimately produce the physics analysis usable representation of the Events, e.g. the four vectors of particle trajectories uncovered in the data. For this configuration, the jobs were writing files with an average Event size on disk of 0.62 MB/Event.

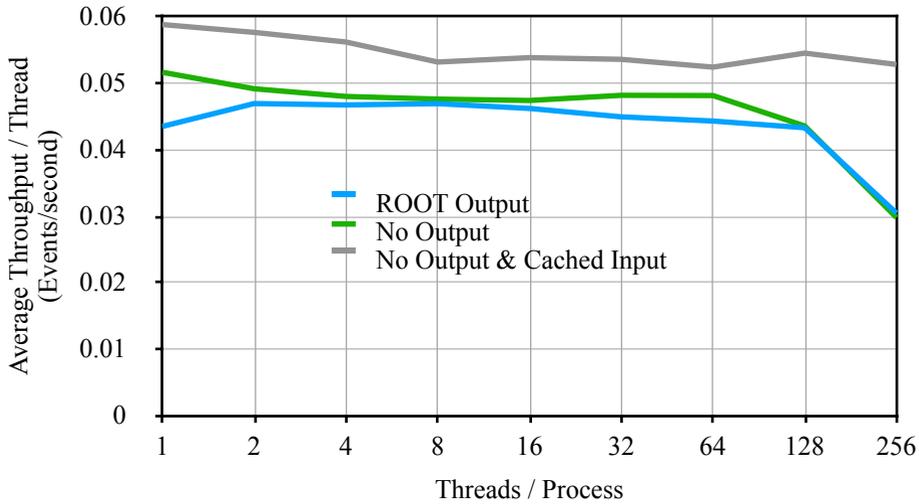

**Fig. 5.** Average event throughput per thread as a function of the number of threads used in the reconstruction process. Rates with and without writing output as well as without output and reading from a memory cache are shown.

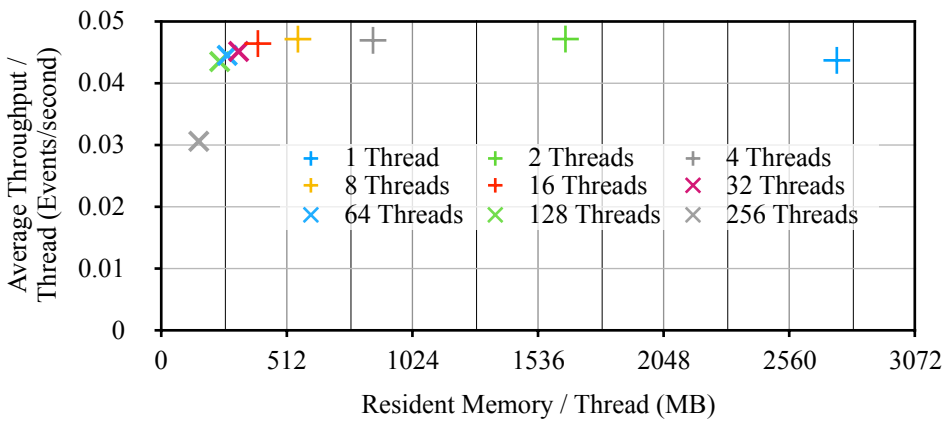

**Fig 6.** Average event throughput per thread as a function of average resident memory used per thread for reconstruction process. Each point represents the measurement of a process using that number of threads.

Figure 5 shows the Event processing rate per thread for this step as a function of the number threads used by a job. From the figure we see that removing the output of the ROOT file did not appreciably change the thread scaling. However, replacing the input with a memory cache allowed perfect thread scaling up to 256 threads. This implies that for this job configuration, input is a more substantial scaling limit than output.

Figure 6 shows the Event processing rate per thread versus the amount of resident memory needed per thread as the number of threads used in a job are varied. This is for the full I/O configuration of the job. The figure shows that CMSSW can run reconstruction below 2GB per core by using more than one thread in the job and the code scales well up to 128 threads where it could fit within an average memory usage of 256MB per thread.

## 4.Conclusion

The design of the CMS software supports excellent Event throughput scaling as the number of threads used in a job are increased. The implementation allows CMSSW workflows to run on machines with less memory than most machines on the US computational grid where those machines typically have 2 GB per core.

The primary scaling limit is from the use of ROOT I/O. The extent to the limitation is driven by the fraction of time spent in the I/O system compared to doing the needed computations. The generator/simulation and reconstruction steps both have large computation compared to the I/O time to have good thread scaling up to the use of 64 threads. In contrast, the overlay with HLT workflow runs at a much higher event throughput and has the largest output rate of any of the workflows. This leads to that workflow losing thread efficiency at around 16 threads.